# Is the unfoldome widespread in proteomes?

Antonio Deiana[1,2]*, Andrea Giansanti[1,2,3]*[§]

[1]Department of Physics, La Sapienza University of Rome, P.le A. Moro 5, 00185, Rome, Italy
[2]Interdepartmental Research Centre for Models and Information Analysis in Biomedical Systems (CISB), La Sapienza University of Rome, C.so Vittorio Emanuele II 244, 00186, Rome, Italy
[3]INFN, Sezione di Roma, P.le A. Moro 2, 00185, Rome, Italy

*These authors contributed equally to this work
[§]Corresponding author

Email addresses:
AD: Antonio.Deiana@roma1.infn.it
AG: Andrea.Giansanti@roma1.infn.it

# Abstract

The term unfoldome has been recently used to indicate the universe of intrinsically disordered proteins. These proteins are characterized by an ensemble of high-flexible interchangeable conformations and therefore they can interact with many targets without requiring pre-existing stereo-chemical complementarity. It has been suggested that intrinsically disordered proteins are frequent in proteomes and disorder is widespread also in structured proteins. However, several studies raise some doubt about these views.

It this paper we estimate the frequency of intrinsically disordered proteins in several living organisms by using the ratio S between the likelihood, for a protein sequence, of being composed mainly by order-promoting or disorder-promoting residues. We scan several proteomes from Archaea, Bacteria and Eukarya. We find the following figures: 1.63% for Archaea, 3.91% for Bacteria, 16.35% for Eukarya. The frequencies we found can be considered an upper bound to the real frequency of intrinsically disordered proteins in proteomes. Our estimates are lower than those previously reported in several studies. A scanning of proteins in the Protein Data Bank (PDB) searching for segments of non-observed residues reveals that segments of non-observed residues longer than 30 amino acids, are rare.

Our observations support the idea that the spread of the unfoldome has been often overestimated. If we exclude some exceptions, the structure-function paradigm is generally valid and pre-existing stereo-chemical complementarity among structures remains an important requisite for interactions between biological macromolecules.

## Background

An important paradigm of structural biology states that proteins fulfil their biological function through a well-defined three-dimensional configuration. This structure-function paradigm is important to explain the high specificity in the interactions among many proteins and substrates, as for example in enzyme catalysis [1-3]. In the classical view, substrates bind to the active site of stereo-chemically compatible proteins. In this way their interaction is characterized by high specificity and affinity. In the last decade it has been discovered a growing number of intrinsically disordered proteins characterized by an ensemble of high-flexible interchangeable conformations [4]. The biological function of intrinsically disordered proteins has been extensively studied [4-9], as well as their role in the development of several diseases [4,8,10-14]. Their discovery suggests that pre-existing stereo-chemical complementarity is not necessary for the interactions among proteins and substrates: there are proteins that assume a three-dimensional structure complementary to the target just *on-fly*, after the interaction has begun. In this way, the interaction is highly specific, but the affinity between protein and substrate is low. Moreover, since disordered proteins are thermodynamically unstable, they quickly change the conformation they get in the interaction, losing the stereo-chemical complementarity with the substrate and unbinding it. Therefore, they have a high turn-over. The considerations above reported suggests that it is necessary to develop new ideas on the way these proteins work.

It is important to note that in the literature the term *disorder* is used to indicate both folded proteins with unstructured domains in their tertiary structure and globally unfolded proteins. To avoid confusion, in this paper we indicate the former proteins as *intrinsically disordered* and the latter as *natively unfolded*. In recent papers the term *unfoldome* has been introduced to indicate the universe of intrinsically disordered proteins [8, 14-16]. An interesting point is the question whether this unfoldome is widespread in proteomes or not. It has been suggested that intrinsically disordered proteins are highly abundant in living organisms [16-19]. However, several studies in the literature are controversial. In this study we revisit this question by estimating both the frequency of disorder in folded proteins and the frequency of natively unfolded ones.

The spread of disorder in structured proteins is an important issue since it gives valuable information on their affinity to a specific target. This issue has been investigated by several authors. They analyse the tertiary structures of the proteins deposited in the Protein Data Bank (PDB) [20]. Le Gall *et al.* consider both non-observed and ambiguous residues [21]. A residue is *non-observed* if its atom spatial coordinates are not experimentally resolved, whereas it is *ambiguous* if its spatial coordinates are resolved in some structures and they are not resolved in other structures sharing the same amino acid sequence. In their study Le Gall *et al.* conclude that about 40% of the proteins in the PDB have domains made of non-observed or ambiguous residues [21]. More recent papers observe that ambiguous residues might be due to experimental or environmental conditions [22,23], therefore their presence is not to be related to intrinsic functional reasons. Ambiguous residues are not considered by Lobanov *et al.* in compiling a library of disordered patterns from proteins in the PDB [24]. It should be interesting to re-estimate the frequency of unstructured domains in folded proteins by considering only non-observed residues in the PDB.

As regards to natively unfolded proteins, their frequency in proteomes has been investigated by means of different computational methods [4, 17-19]. These studies are important since they help in understanding how low affinity interactions are widespread both in healthy and pathological cellular processes. Unfortunately, the published results are highly dependent on the method adopted. Ward *et al.* use the DISOPRED predictor to estimate the percentage of natively unfolded proteins in proteomes [18]. They report the following figures: 2% in Archaea, 4.2% in Bacteria, 33% in Eukarya. These estimates are quite different from those reported by Dunker *et al.* by using PONDR [17]. In some cases, the differences are really surprising. As an example, the frequency of natively unfolded proteins in Drosophila Melanogaster is estimated as some 36.6% by DISOPRED and about 63% by PONDR [4, 17, 18]. These discrepancies deserve a discussion and they raise the question about the real spread of the unfoldome in proteomes.

In this study we re-evaluate the occurrence of disorder in folded proteins by looking at the protein structures deposited in the PDB. With disorder we operationally indicate the presence of residues missing spatial coordinates in the tertiary structures deposited in the PDB [18, 25]. A residue therefore is disordered if its spatial coordinates are missing in the experimentally resolved tertiary structure. In the following we refer to disordered residues as *non-observed*. Consequently, a polypeptide chain segment is disordered if it is made by consecutive non-observed residues. We consider 133710 protein chains, including complexed proteins, without considering the structural or functional class they belong to. We do not consider ambiguous residues and experimental conditions. In this way we estimate the greatest amount of disorder that it is possible to extract from known folds. We find 69% of the structures in the PDB with at least one non-observed residue; of these proteins, 27% have segments longer than 10 non-observed residues and 5% have segments longer than 30 residues. Therefore, the frequency of proteins with long disordered segments is low. A quick check has revealed that 51% of the segments of non-observed residues found are made by less than 5 amino acids, indicating that they might be noise of the experimental procedure used to resolve the structure. The other 49% of the segments found are generally of short length and they involve less than 20% of the amino acids. From these observations we conclude that disorder in structured proteins is rare and it hardly implies low affinity interactions with substrates. Further studies however are necessary to check whether the presence of these segments is due to intrinsic propensities of the amino acid sequence to have unstructured domains of possible functional importance or to an intrinsic difficulty to crystallize.

It is also probable that the frequency of natively unfolded proteins in living organisms has been overestimated in some works. As said above, up to 30% of the proteins in Eukaryotic organisms have been reported as natively unfolded. More precisely, Ward *et al.* consider amino acid sequences with segments predicted as disordered longer than 30 residues [18]. The estimate they report therefore is significantly higher than the frequency of structured proteins with long segments of non-observed residues that is about 5%. Orengo *et al.* [26] report that about 80% - 90% of the amino acid sequences in living organisms can be associated to a known fold. The rationale behind their analysis is the observation that many amino acid sequences share the same fold. To obtain their result, they align protein sequences through Hidden Markov Model (HMM). A direct consequence of their result is the following: only 10% - 20% of amino acid sequences in proteomes cannot be associated to a known fold. If we consider that about all folds have been already discovered [27], we conclude that natively unfolded proteins must not exceed 20% of the sequences in proteomes, so the frequency of natively unfolded proteins reported in the literature seems to be overestimated. In a previous work we have shown that different predictors of natively unfolded proteins generally do not agree in classifying as folded or unfolded a certain number of amino acid sequences; the percentage of these "ambiguous" proteins depends on the dataset and it ranges from 10% up to 30% [28]. By using a consensus score $S_{SU}$ among three predictors of natively unfolded proteins, namely Poodle-W [29], gVSL2 [28, 30, 31] and mean pairwise energy [32], we have shown that the "ambiguous" proteins belong to a twilight zone between order and disorder, in the amino acid compositional space [28]. The structural properties of proteins in the twilight zone have not been yet extensively investigated. However, we have shown that they do not exhibit a tendency to have long flexible segments or loops in their tertiary structure (see figure 6 in [28]).

It has been reported that folded polypeptide chains are enriched in T, C, F, I, Y, V and L, while unfolded one are enriched in M, A, R, Q, S, P, E and K. The first type of amino acids have been named order-promoting, the second one disorder-promoting [33]. Interestingly, we have shown in [26] that proteins classified as unfolded by $S_{SU}$ are enriched in disorder-promoting amino acids. We evaluated the ratio S between the likelihoods, for a sequence, of being composed mainly by order-promoting and disorder-promoting amino acids [34]. Proteins classified as unfolded by $S_{SU}$ generally have a negative S score. In this paper we use the S score to separate amino acid sequences enriched in order-promoting residues (S > 3.24) from those enriched in disorder-promoting residues (S < -0.83), the latter being candidates to be natively unfolded [28]. Proteins with S scores between -0.83 and 3.24 are considered to belong to the twilight zone. We verify that proteins with S scores lower than -0.83 have a higher probability that their non-observed residues make long segments in the polypeptide chain. These observations suggest that proteins with long non-observed segments in the polypeptide chain must be searched for among those with S scores lower than -0.83. However, many proteins in the PDB with negative S scores do not have a high frequency of non-observed residues in their tertiary structure. This result indicates that amino acid composition is not sufficient to infer structural properties of proteins. The frequency of proteins with S scores lower than -0.83 can be considered therefore as an upper bound to the real frequency of natively unfolded proteins. The application of S to several proteomes from Archaea, Bacteria and Eukarya has given the following figures: 1.63% for Archaea, 3.91% for Bacteria, 16.35% for Eukarya. As expected, our estimates of the frequency of natively unfolded proteins are lower than those previously reported. Interestingly, our estimates are consistent with the observation by Orengo *et al.* [26]. The consistence of our estimates with those by Orengo *et al.* suggests that the spread of the unfoldome has been often overestimated in several works and it should not exceed 16% of proteins in living organisms. Natively unfolded proteins seem to be exceptions in the universe of functional protein sequences and pre-existing stereo-chemical complementarity appears to be important for a protein to fulfil a cellular function.

## Results

### Analysis of disorder in the folded proteins deposited in the PDB

In this paper we evaluate the frequency of disorder in the proteins deposited in the PDB by considering non-observed residues in their tertiary structure. Non-observed residues in a protein tertiary structure are those missing atom spatial coordinates. To find these residues we align the protein sequence as extracted from the SEQRES fields and from the ATOM fields of the PDB files. Residues that are present in SEQRES fields and that are not present in the ATOM fields are considered non-observed [18, 25].

In figure 1 we plot the distribution of non-observed residues in the PDB. We observe that 43% of the proteins do not have non-observed residues, whereas the fraction of non-observed residues is below 10% of all residues in the protein sequences in 47% of the proteins analysed, it is between 10% and 20% in 7% of the proteins analysed and it is above 20% in only 1% of the proteins in the PDB. We conclude that, generally, in the structures deposited in the PDB, the frequency of non-observed residues is low. In table 1 we report the fraction of proteins with segments of non-observed residues shorter than 10 amino acids, between 10 and 30 amino acids and longer than 30 amino acids, respectively. As we can see, 69% of the proteins considered have segments shorter than 10 non-observed residues, 27% have segments between 10 and 30 residues and only 5% have segments longer than 30 non-observed residues. A more detailed statistics reveals that 46% of the proteins have segments shorter than 5 residues and 29% have segments between 5 and 10 residues. From these results we conclude that the frequency of structured proteins with long disordered segments is low.

We now investigate about the length of segments of non-observed residues and their relationship with the length of the amino acid sequence. In figure 2 we plot the frequency of segments of non-observed residues as a function of their length. We find a distribution similar to that reported by Lobanov *et al.*[24]. About 15% of non-observed residues are singletons, and 30% makes segments shorter that 4 amino acids. There is consensus in the idea that segments of non-observed residues shorter than 4 amino acids might be due to noise in the experimental procedures used to resolve the tertiary structures, so these segments do not indicate the presence of disorder in proteins. On the other hand, 55% of the segments that we observe can indicate unstructured loops or domains in the protein structures.

In figure 3 we plot the distribution of the ratio between the length of segments made of non-observed residues and the length of the amino acid sequence. We see that the distribution rapidly decreases, indicating that generally segments of non-observed residues involve a low fraction of the amino acid sequence. More precisely, 64% of the segments found involve a percentage lower than 3% of the amino acids that make the polypeptide chain, 28% of the segments involve a percentage of amino acids between 4% and 10%, 7% of the segments involve a percentage between 11% and 20% and only 1% of the segments involve more than 21% of the amino acids that make the protein sequence.

From the above results we conclude that non-observed residues generally are singletons or they make short segments in the tertiary structure of proteins, involving a low fraction of the polypeptide chains. Long disordered segments, longer than 30 residues, are rare.

## Analysis of disorder in proteomes

The frequency of natively unfolded proteins in living organisms can be estimated by scanning proteomes with predictors of global disorder. As said above in the Background, the estimated percentage of natively unfolded proteins is highly dependent on the predictor used. Ward *et al.* used DISOPRED to scan several organisms from Archaea, Bacteria and Eukarya, searching for sequences with segments predicted as disordered longer than 30 residues [18]. They report the following figures: 2% in Archaea, 4.2% in Bacteria, 33% in Eukarya. In a previous work we considered three predictors of natively unfolded proteins: Poodle-W, gVSL2 and mean pairwise energy [28]. In this paper we re-evaluated the percentage of natively unfolded proteins in living organisms by using both gVSL2 and mean pairwise energy. By using gVSL2 we obtained the following figures: Archaea 5.45%, Bacteria 11.11%, Eukarya 36.15%. By using mean pairwise energy, we obtained the following figures: Archaea 1.53%, Bacteria 10.48%, Eukarya 28.20%. The complete results are reported in the Appendix. As we can see, the percentage of natively unfolded proteins is highly dependent on the method used. gVSL2 returns percentage significantly higher than those obtained through mean pairwise energy.

In an important study Orengo *et al.* show that the percentage of protein sequences in proteomes that cannot be associated to a known fold is 10% - 20% [26]. This observation raises the question whether the frequencies of natively unfolded proteins evaluated by disorder predictors are not overestimated. In a previous work we used a consensus score $S_{SU}$ among Poodle-W, gVSL2 and mean pairwise energy to analyse proteins in datasets [28]. We have found that a percentage from 10% to 30% of sequences belong to a twilight zone between order and disorder, in the amino acid compositional space. These proteins are classified by some predictors as folded and by other predictors as unfolded; therefore they have an amino acid composition that makes hard an unambiguous classification in the folded or unfolded class. In [28] we have shown that proteins in the twilight zone do not show a tendency to have a high frequency of non-observed residues (see figure 6 in [28]).

Proteins classified as unfolded by $S_{SU}$ are enriched in disorder-promoting amino acids. In [28] we have characterized the bias in the amino acid composition by using the ratio S between the likelihood, for a sequence, of being composed mainly by order-promoting and disorder-promoting residues. The score S is positive (negative) if the protein is enriched in order-promoting (disorder-promoting) amino acids (see Methods). We have shown that proteins belonging to the twilight zone have an S score around 0, indicating a balanced mixture of order-promoting and disorder-promoting residues. In this work we consider all protein sequences with an S score between -0.83 and 3.24 as belonging to the twilight zone. The thresholds of the S score have been determined so to include 90% of the proteins classified in the twilight zone by $S_{SU}$ (see Methods for details). In figure 4 we report the logarithm plot of the probability that non-observed residues make segments of a given length as a function of the lengths of the segments. We observe that the fraction of non-observed residues scales as -1.31 ± 0.11 in proteins enriched in order-promoting residues, as -0.58 ± 0.14 in proteins enriched in disorder-promoting residues and as -0.92 ± 0.13 in proteins belonging to the twilight zone. This result shows that, in proteins enriched in disorder-promoting residues, non-observed amino acids have the highest probability to make long segments in the polypeptide chain. Therefore, proteins candidates to be natively unfolded must be searched for among those classified as unfolded by $S_{SU}$ or, alternatively, among those with negative S scores, lower than -0.83.

It is important to note however that only a fraction of the proteins with negative S scores have a high frequency of non-observed residues. In figure 5 we plot the distributions of these residues in proteins with S scores lower than -0.83, between -0.83 and 3.24 and above 3.24, respectively. It is evident that a high frequency of proteins with negative S scores do not have non-observed residues in their structures. This result shows that amino acid composition is not sufficient to infer whether a protein sequence has non-observed residues in their tertiary structure. Amino acid composition, therefore, is only a necessary condition to have long non-observed segments in the polypeptide chain, but it is not a sufficient one. Proteins with negative S scores do not necessarily have long non-observed residues in the tertiary structure, since other factors affect the stabilization of protein domains, as the order of the amino acid in the primary structure and environmental conditions. On the other hand, the observation that proteins with long non-observed segments of the polypeptide chain have a high probability to have negative S scores suggests that the frequency of natively unfolded proteins obtained through S can be consider an upper bound to the real frequency of these proteins in proteomes.

In table 2 we report the frequency of natively unfolded proteins obtained by scanning with S several proteomes from Archaea, Bacteria and Eukarya. We find the following figures: 1.63% in Archaea, 3.91% in Bacteria, 16.35% in Eukarya. Interestingly, the frequency of natively unfolded proteins that we estimate with the S score is lower than those previously reported and consistent with the finding by Orengo *et al.* [26].

## Discussion

In several papers it has been suggested that intrinsically disordered proteins are abundant in living organisms [4, 8, 17-19]. The existence of these proteins shifts the classical structure-function paradigm that postulates a strict relation between the tertiary structure of a protein and its biological function. High-flexible domains in protein structures imply that proteins and substrates can bind through dynamics that do not require pre-existing stereo-chemical complementarity. In this way the interaction between a protein and a substrate is characterized by a high specificity, but a low affinity and a rapid turn-over. This kind of interactions can be an advantage in signalling and cell regulation processes, since they allow to proteins to bind different targets and therefore to trigger different processes [4-9]. However, in some cases, they can facilitate the onset of diseases through the binding to erroneous targets or through the aggregation in amyloidal fibrils [10-16]. An estimate of the spread of unfoldome is of importance, since it gives valuable information about how protein interaction networks of organisms are organized

It has been reported that the frequency of natively unfolded proteins is higher in Eukarya than in Archaea and Bacteria [17-19]. As an example, Ward *et al.* report 2% of the proteins in Archaea, 4.2% in Bacteria and 33% of the proteins in Eukarya are natively unfolded [18]. The higher frequency of natively unfolded proteins in Eukarya has been related to the tendency of these proteins to be involved in cellular regulation and to the more complex regulatory processes typical of Eukaryotic organisms [18]. The estimated frequency is highly dependent on the computational method used. In some cases the discrepancies are really significant. As an example, the estimated frequency in Drosophila Melanogaster is 33% by using DISOPRED and 62% by using PONDR [4, 17, 18]. Disorder predictors consider mainly amino acid composition to infer whether a protein sequence folds into a tertiary structure. In this paper we have shown that amino acid composition is not sufficient to determine whether a protein has disordered segments in their polypeptide chains, confirming suggestion by several papers [35]. This fact can explain the discrepancies observed in evaluating the frequency of natively unfolded proteins. We have shown that the score S is effective in selecting out from dataset sequences enriched in disorder-promoting amino acids, and these sequences have a higher probability that their non-observed residues make long segments in the polypeptide chains. Therefore, we think that the frequency of natively unfolded proteins estimated through the S score can be considered as an upper bound to the real frequency of natively unfolded proteins in proteomes. We find that 1.63% of proteins in Archaea have amino acid composition typical of unfolded ones, 3.91% in Bacteria and 16.35% in Eukarya. In our opinion the real frequency of natively unfolded proteins in living organisms does not exceed these figures and therefore it is lower than those previously suggested in the literature. Interestingly, our estimates are consistent with the results reported by Orengo *et al.* [26]. In their work they observe that a large number of the protein sequences in proteomes share the same fold. By aligning amino acid sequences with Hidden Markov Model, they estimate that only a percentage from 10% to 20% of sequences in proteomes cannot be associated to a known fold. These sequences are often singletons, and they suggest that natively unfolded proteins must be searched for among them. It is interesting that we obtain similar estimates for natively unfolded proteins in Eukarya, about 16%, lower than the percentage of 33% previously reported. It will be interesting to re-estimate the frequency of proteins enriched in disorder-promoting amino acids in protein interaction networks and in proteins involved in the development of diseases.

The frequency of disorder has been probably overestimated also in structured proteins. A significant amount of unstructured domains in folded proteins can indicate that sequences with a tertiary structure can bind targets with low affinity, as done natively unfolded proteins. Therefore, an estimate of the amount of disorder in folded proteins is an important issue. Le Gall *et al.* analyse the tertiary structures of the proteins deposited in the PDB searching for non-observed and ambiguous residues [21]. The latter are identified by considering the redundancy in the PDB: residues are considered ambiguous if they are observed in some structures but they are non-observed in other structures sharing the same amino acid sequence. On the other hand, a residue is non-observed if it is impossible to determine its spatial position in all structures with the same amino acid sequence. They report that about 10% of the proteins in the PDB contain segments of the polypeptide chain longer than 30 residues ambiguous or non-observed, and the percentage arises to 40% if segments between 10 and 30 amino acids are considered. From these observations it is possible to speculate that disordered domains are frequent in the tertiary structures of proteins and it may influence their dynamics. The existence of ambiguous residues has been studied also by Zhang *et al.* [23]. They observe that segments of the polypeptide chain made of ambiguous residues are *dual personality fragments* that can be ordered or disordered depending on the environmental conditions. They suggest that this dual personality behaviour is due to a peculiar amino acid composition and it gives to proteins the ability to modify their dynamics in response to a change in the environment. Mohan *et al.* [22] report that the identification of non-observed residues in proteins is dependent from the experimental conditions used to resolve the structure, like temperature, pH, salt concentration and time of crystallization. They analyse protein structures with sequence identity higher than 90% and resolved in different experimental conditions and they observe that in many cases non-observed residues do not coincide in the structures analysed. These observations therefore suggest that the presence of ambiguous residues is due to environmental conditions that can affect the structure of the proteins. In our opinion these two studies are not sufficient to conclude whether the presence of ambiguous residues in a protein structure is an intrinsic property written in its amino acid sequence of it is an effect of environmental conditions related to the experimental set-up. It is well-known that the presence of osmolytes in solution can shift an amino acid sequence towards the folded or the unfolded state by modulating the number of hydrogen-bonds in the backbone of the polypeptide chains, and this effect does not depend from amino acid composition of the protein sequence [36]. In any case, it seems reasonable to conclude that the presence of ambiguous residues does not necessarily imply that a domain in a protein fold has an intrinsic propensity to remain unstructured. Ambiguous residues have not been considered by Lobanov *et al.* in compiling a library of disordered patterns from proteins in the PDB [24].

In the present study we re-evaluate the frequency of disorder in the proteins of the PDB by considering only non-observed residues in the tertiary structures. We consider a residue as non-observed if its spatial coordinates are missing in the tertiary structure experimentally resolved. Differently by Le Gall *et al.* [21], therefore, we do not include in the number of disordered proteins those with *ambiguous observed* residues. We find that 69% of the proteins in the PDB have non-observed residues, but generally they are singletons or they make short segments of the polypeptide chain, of the order of 5 amino acids, while only in 5% of the proteins they make segments longer than 30 residues and in 27% of the proteins they make segments between 10 and 30 residues. Our estimates therefore indicate a frequency of proteins with disordered segments lower than the previously reported one and they suggest that disorder is not frequent in structured proteins. A quick analysis has shown that 45% of the disordered segments found are singletons or they are made by less than 4 residues, so they probably do not indicate intrinsically unstructured domains in the protein fold but they might be due to experimental noise in the characterization of the structure. The remaining 55% of them are mainly of short length. An analysis of the relation of the length of the segments of non-observed residues and the length of the amino acid sequence has revealed that these segments generally involve a low fraction of the polypeptide chains, and there is no evidence that their length is dependent from the length of the protein sequence. The frequency of proteins with long disordered segments that involve more than 20% of the polypeptide chain is not above 1% of the structured proteins currently known.

From the experimental data currently at disposal we conclude that in the structured proteins disordered segments involve generally short regions of the polypeptide chains and therefore they hardly affect significantly protein dynamics or imply low affinity interactions with substrates. Further studies however are necessary to clear up whether the presence of long disordered segments in these proteins are due to an intrinsic propensity to have unstructured high-flexible loops of possible functional importance or to an intrinsic difficulty to crystallize.

In conclusion we think that the spread of the unfoldome has been overestimated in many works. There is no doubt that there exist high-flexible proteins that interact with substrates without requiring pre-existing stereo-chemical complementarity. However, unstructured domains are not frequent in the known structured proteins, and the frequency of natively unfolded proteins in proteomes probably does not exceed 16%. These results point to a general validity of the structure-function paradigm and pre-existing stereo-chemical complementarity is generally important for the interactions among biological macromolecules.

# Methods

## Datasets

To study the frequency of disorder in folded proteins, we extensively analysed all protein structures deposited in the Protein Data Bank (PDB) database, version 23, November, 2010 [20]. In the total we analysed 63590 proteins.

To delimit the twilight zone in the S space we consider the set C used in [28] to test $S_{SU}$. Folded proteins are extracted from PDBSelect25 [37, 38], version October 2007, which contains 3694 proteins with sequence identity lower than 25%. Structures with a resolution above 2 Å and an R-factor above 20% are excluded. A restricted list of 2369 is obtained, 1573 of which with a percentage of disordered amino acids below 5%. Operationally, a residue is considered as disordered if it is present in the SEQRES but not in the ATOM field of the PDB file [18, 25]. A list of 81 natively unfolded proteins, with at least 70% of disordered amino acids and sequence identity below 25%, are extracted from the DisProt database [39], version 3.6.

## Strictly unanimous consensus score among predictors of natively unfolded proteins

The strictly unanimous consensus score $S_{SU}$ is a consensus index among different predictors of natively unfolded proteins. It combines three predictors: mean pairwise energy [32], *gVSL2* [30, 31], and Poodle-W [29]. The protocol used to compute these indexes is described in the previous chapters [28]. Mean pairwise energy is the arithmetic mean of the global pairwise energy of the protein sequence, and therefore it is an estimate of the pairwise energy per residue [32]. *gVSL2* is the arithmetic mean of the disorder scores obtained through *VSL2* [30, 31], a good performing disorder predictor. Poodle-W [29] evaluates whether a protein sequence is disordered through a spectral graph transducer [40].

The strictly unanimous consensus score $S_{SU}$ [28] considers the three predictions by mean pairwise energy, *gVSL2* and Poodle-W. If the three predictors agree in classifying a protein as folded, $S_{SU}$ classifies the protein as ordered. If the three predictors agree in classifying a protein as unfolded, $S_{SU}$ classifies the protein as disordered. If at least two predictors disagree in classifying a protein as folded or unfolded, $S_{SU}$ assigns the protein to the twilight zone.

## Log-odds ratio of the likelihoods that a sequence has amino acidic composition typical of folded and unfolded proteins

Referring to a simple probabilistic model, one assumes to have reliable estimates of the probability of occurrence of each amino acid in folded and unfolded proteins $\{\pi_a^{(F)}\}$ and $\{\pi_a^{(U)}\}$, respectively where a runs over all amino acid labels. We estimated these probabilities on the set of folded and natively unfolded proteins selected by Shimizu et al. to test Poodle-W [109]. Then, a folded protein can be thought of as if its sequence were sampled from $\{\pi_a^{(F)}\}$ through a sequence of independent extractions. The likelihood that a sequence has amino acidic composition typical of a folded protein is:

$$L_F = \prod_{a=1}^{20} \left(\pi_a^{(F)}\right)^{n_a}$$

where $\pi_a^{(F)}$ is the probability of amino acid *a* and $n_a$ is the occurrence of amino acid *a* in the sequence. The probabilistic model implicit in the above definition is a 0-order Markov chain. Similarly we can define $L_U$ by using $\pi_a^{(U)}$. $L_F/L_U$ is the ratio of the likelihoods, for a given sequence and through its amino acidic composition $\{n_a\}$, to have been generated from $\{\pi_a^{(F)}\}$ and $\{\pi_a^{(U)}\}$, respectively. The log-odds ratio of a given sequence is then defined as:

$$S = \sum_{a=1}^{20} n_a \cdot \ln\left(\frac{\pi_a^{(F)}}{\pi_a^{(U)}}\right)$$

Order-promoting amino (disorder-promoting) acids contribute with positive (negative) terms to S, since their ratios $\pi_a^{(F)}/\pi_a^{(U)}$ are bigger (lower) than one. Therefore, S is positive (negative) if the protein is composed mainly by order-promoting (disorder-promoting) amino acids. When a protein is composed by approximately the same number of order-promoting and disorder-promoting amino acids, its S score has a value close to zero.

## Definition of the twilight zone in the S space

To define the twilight zone in the S space, we use the following procedure. We compute the S score of the 2369 proteins selected in [28] to test $S_{SU}$. For these proteins, we have the list of those classified by $S_{SU}$ as folded, unfolded and in the twilight zone, respectively.

We consider in first instance proteins classified by $S_{SU}$ as folded and those classified in the twilight zone. We plot the histogram of the S scores for these two groups of sequences and then we search for a discriminating line so to identify more than 90% of the proteins classified by $S_{SU}$ in the twilight zone, with the lowest level of false positives. We set the threshold to 3.24.

In second instance we consider proteins classified by $S_{SU}$ as unfolded and in the twilight zone and we repeat the procedure sketched above. We set the threshold to -0.83.

We consider in the twilight zone all protein sequences with S scores between -0.83 and 3.24.

## Probability that a non-observed residue is found in a segment of length *l*

The probability that a non-observed residue is found in a segment of length $l$ is computed by the following algorithm.

Let be $n_s(l)$ the number of segments made by $l$ non-observed residues. Non-observed residue are identified by aligning the protein sequences extracted from SEQRES and ATOM fields of PDB files. Each residue that is present in the SEQRES fields but there are not its spatial coordinates in the ATOM fields is considered as *non-observed*. Let be $n_d$ the total number of non-observed residues in the protein structures deposited in the PDB. The probability that a non-observed residue is found in a segment of length $l$ is given by:

$$p_s(l) = \frac{n_s(l) \cdot l}{n_d}$$

## Tables

### Table 1 - Frequency of proteins with segments of non-observed residues of a given length

| Length of segments | Frequency of proteins |
| --- | --- |
| From 1 to 10 residues | 59% |
| From 11 to 30 residues | 27% |
| Above 30 residues | 5% |

## Table 2 - Frequency of proteins enriched in disorder-promoting amino acids in several proteomes

Proteomes have been download from UniProt:
http://www.uniprot.org/taxonomy/complete-proteomes

| ORGANISM | N. proteins | S < -0.81 | |
|---|---|---|---|
| | | N. | % |
| *Aeropyrum pernix* | 1773 | 37 | 2.81 |
| *Archaeoglobus fulgidus* | 2409 | 43 | 1.78 |
| *Methanococcus jannaschii* | 1771 | 24 | 1.36 |
| *Pyrococcus abyssii* | 1786 | 27 | 1.15 |
| *Thermoplasma volcanium* | 1524 | 20 | 1.31 |
| | **9263** | **151** | **1.63** |
| *Agrobacterium tumefaciens C58* | 5358 | 200 | 3.73 |
| *Aquifex aeolicus* | 1554 | 45 | 2.90 |
| *Chlamydia pneumoniae* | 2739 | 145 | 5.29 |
| *Chlorobium tepidum TLS* | 2281 | 81 | 3.56 |
| *Escherichia coli K12* | 5270 | 129 | 2.45 |
| *Haemophilus influenzae Rd* | 31792 | 835 | 2.62 |
| *Mycobacterium tuberculosis H37Rv* | 60323 | 3009 | 4.99 |
| *Neisseria meningitidis MC58* | 1900 | 67 | 3.53 |
| *Salmonella typhi* | 5350 | 143 | 2.67 |
| *Staphylococcus aureus COL* | 2680 | 88 | 3.28 |
| *Synechocystis species PCC 6803* | 3530 | 116 | 3.29 |
| *Thermotoga maritima* | 1887 | 44 | 2.33 |
| *Treponema pallidum* | 3438 | 110 | 3.20 |
| | **128102** | **5012** | **3.91** |
| *Anopheles gambiae* | 14833 | 1936 | 13.05 |
| *Arabidopsis thaliana* | 50056 | 5741 | 11.47 |
| *Bos taurus* | 15318 | 2554 | 16.67 |
| *Caernorhabditis elegans* | 23353 | 2959 | 12.67 |
| *Drosophila melanogaster* | 33213 | 6198 | 18.66 |
| *Gallus gallus* | 8158 | 1295 | 15.87 |
| *Homo sapiens* | 94909 | 16978 | 17.89 |
| *Mus musculus* | 64890 | 12629 | 19.46 |
| *Oryza sativa* | 97933 | 19063 | 19.47 |
| *Plasmodium falciparum* | 5349 | 142 | 2.65 |
| *Saccharomyces cerevisiae* | 41069 | 3951 | 9.62 |
| | **449081** | **73446** | **16.35** |

# Figures

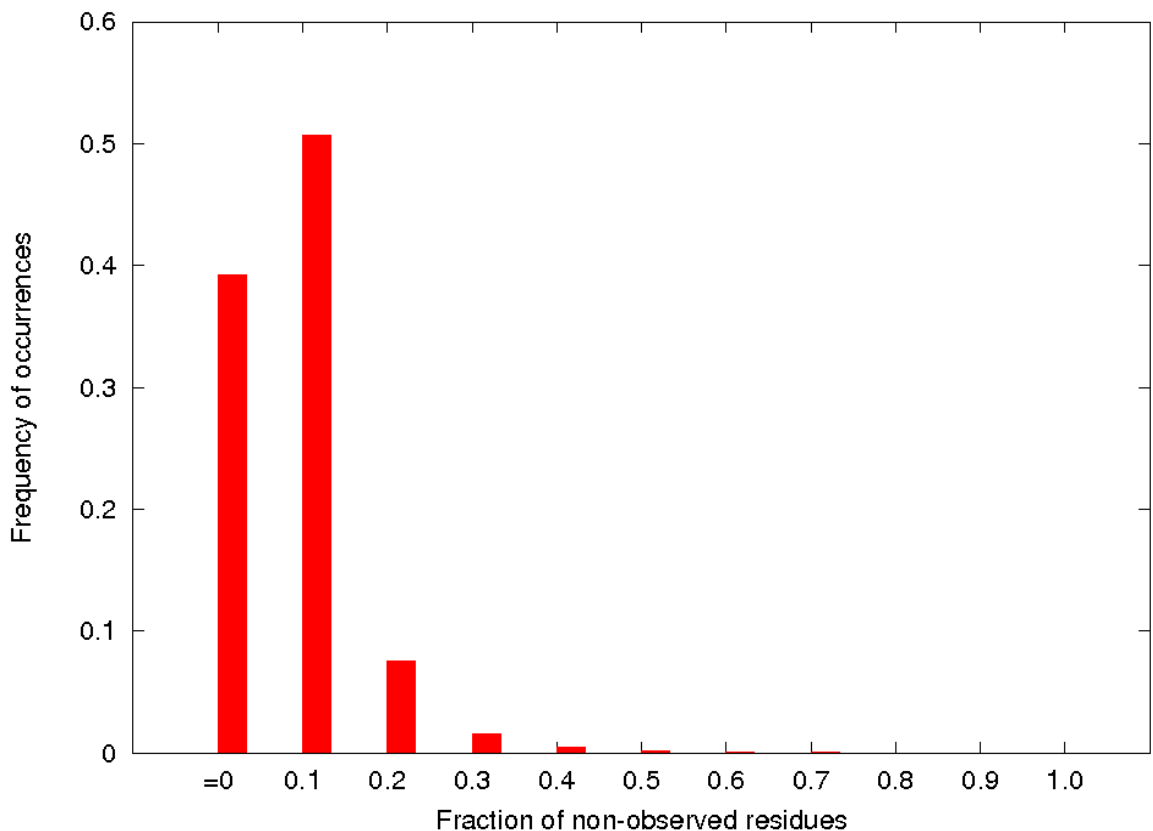

**Figure 1 - Distribution of non-observed residues in the tertiary structures of proteins deposited in the PDB**

The fraction of non-observed residues (horizontal axis) is the ratio between the number of non-observed residues and the number of the amino acids in a protein sequence. The fraction of non-observed residues has been divided in bins. The first bin refers to proteins without non-observed residues. The second bin refers to proteins with a fraction of non-observed residues between 0 and 10%, the third bin refers to proteins with a fraction of non-observed residues between 11% and 20% and so on. The frequency of occurrences (vertical axis) is the number of proteins with a given fraction of non-observed residues divided by the number of proteins considered. 43% of the proteins considered do not have non-observed residues. The frequency of proteins rapidly decreases as the fraction of non-observed residues in the polypeptide chain increases.

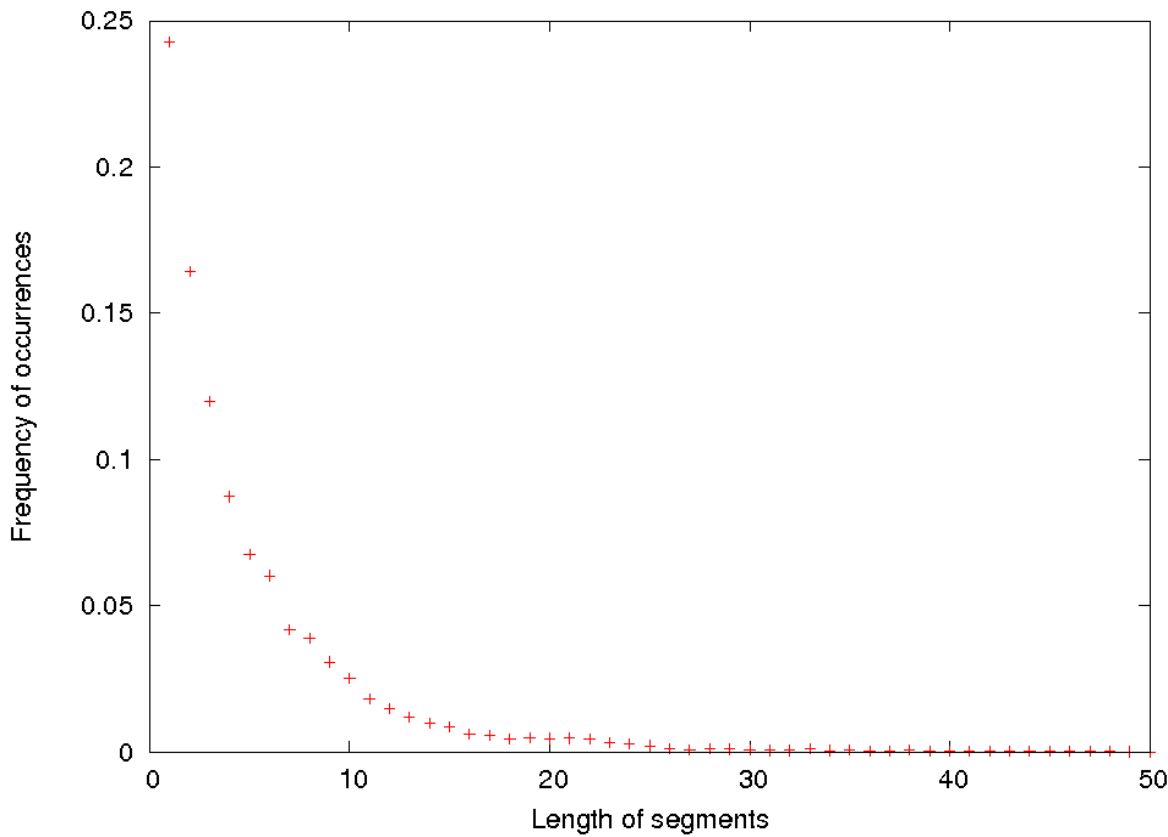

**Figure 2 – Frequency of segments of non-observed residues as a function of the length of the segments**

In the horizontal axis it is reported the length $l$ of a segment made by non-observed residues. In the vertical axis it is reported the number of segments with length $l$ divided by the number of all segments found. 18% of non-observed residues are singletons. 53% of segments found are shorter than 5 amino acids. The frequency of segments rapidly decreases as segment length increases.

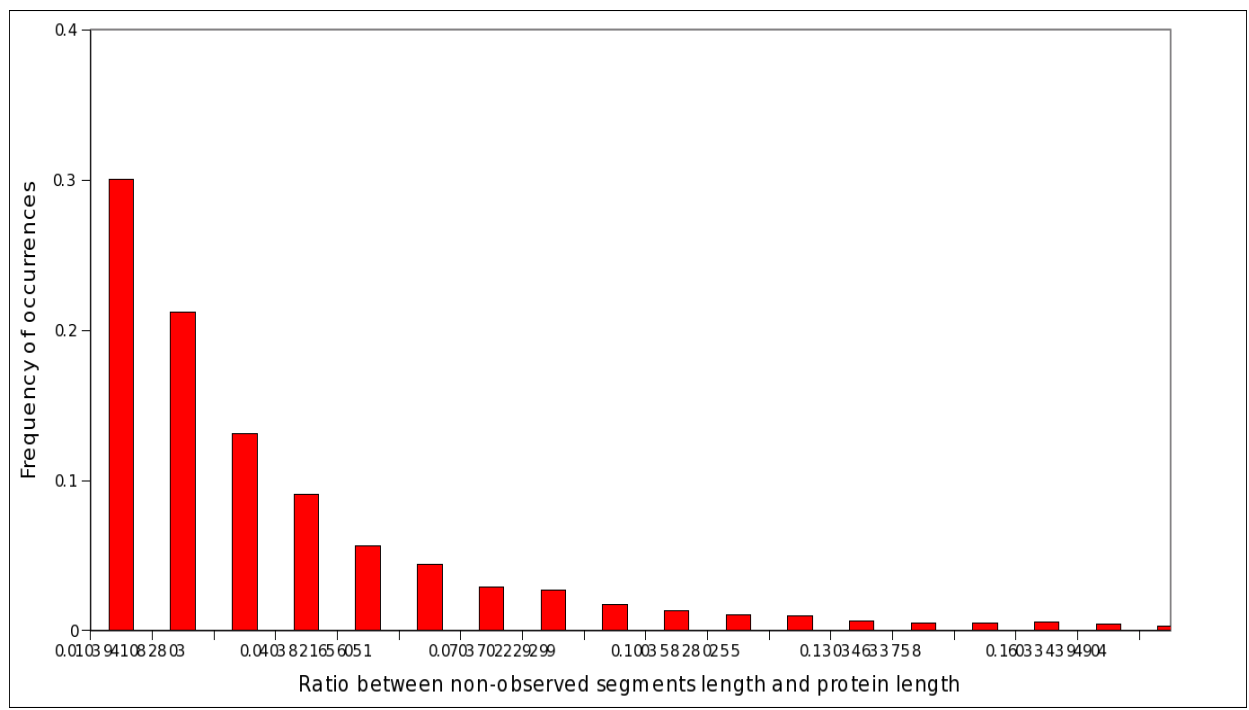

**Figure 3 – Distribution of the ratio between the length of segments of non-observed residues and the length of the amino acid sequence**
In the horizontal axis it is reported the ratio between the length of segments made of non-observed residues and the length of the amino acid sequences. In the vertical axis it is reported the frequency of these ratios in the structured proteins analysed. The distribution rapidly decreases. Segments of non-observed residues tend to involve short fraction of the polypeptide chains in the known structured proteins.

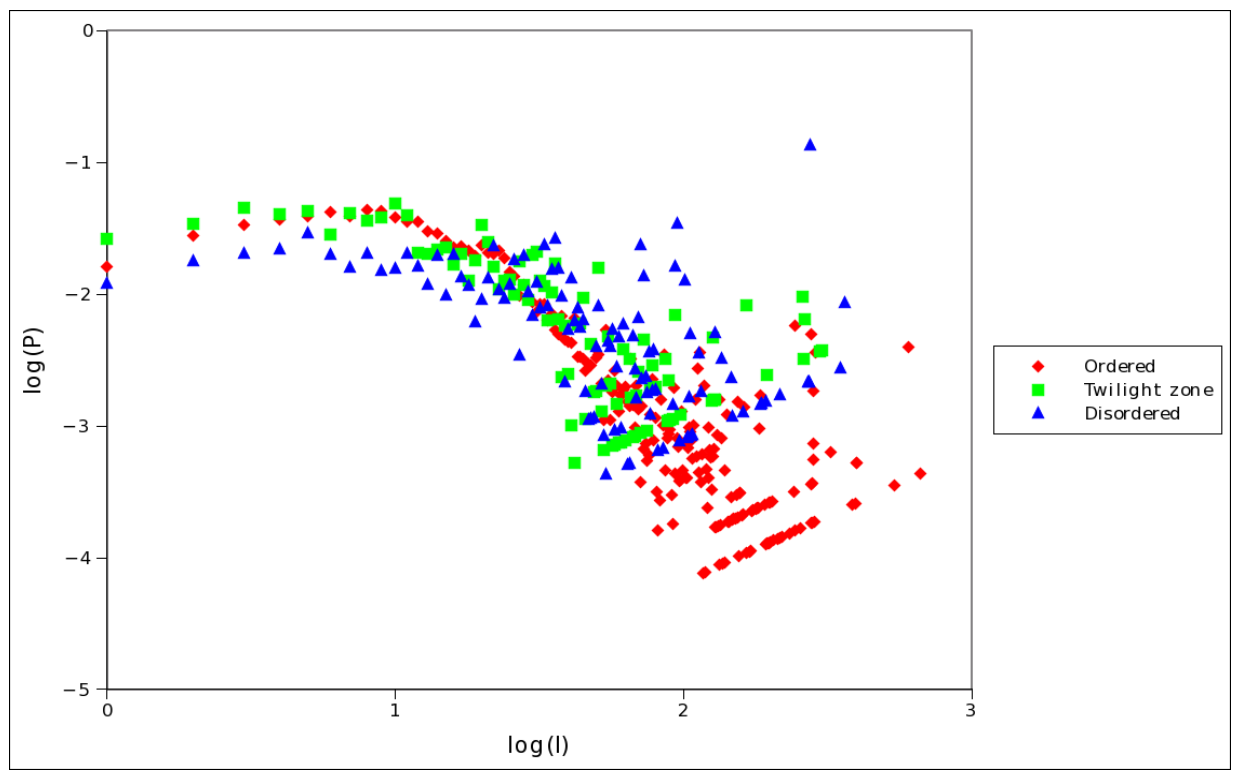

**Figure 4 – Logarithm plot of the probability that a non-observed residue is found in a segment of length *l* in proteins predicted as ordered and disordered and in the twilight zone**

In the horizontal axis it is reported the length $l$ of a segment made by non-observed residues. In the vertical axis it is reported the fraction $n_s(l) \cdot l / n_d$, where $n_s(l)$ is the number of segments of length $l$ and $n_d$ is the number of non-observed residues in the dataset.. Proteins with S > 3.24 (ordered) are plotted in red, proteins with S < -0.83 (disordered) are plotted in blue, proteins belonging to the twilight zone are plotted in green. Proteins predicted disordered by $S_{SU}$ exhibit the higher fraction of non-observed residues involved in segments longer than 30 residues. Fraction of residues involved in segments of length $l$ scales as -1.31 ± 0.11 in proteins predicted ordered, as -0.92 ± 0.13 in proteins belonging to the twilight zone and as -0.58 ± 0.14 in proteins predicted disordered.

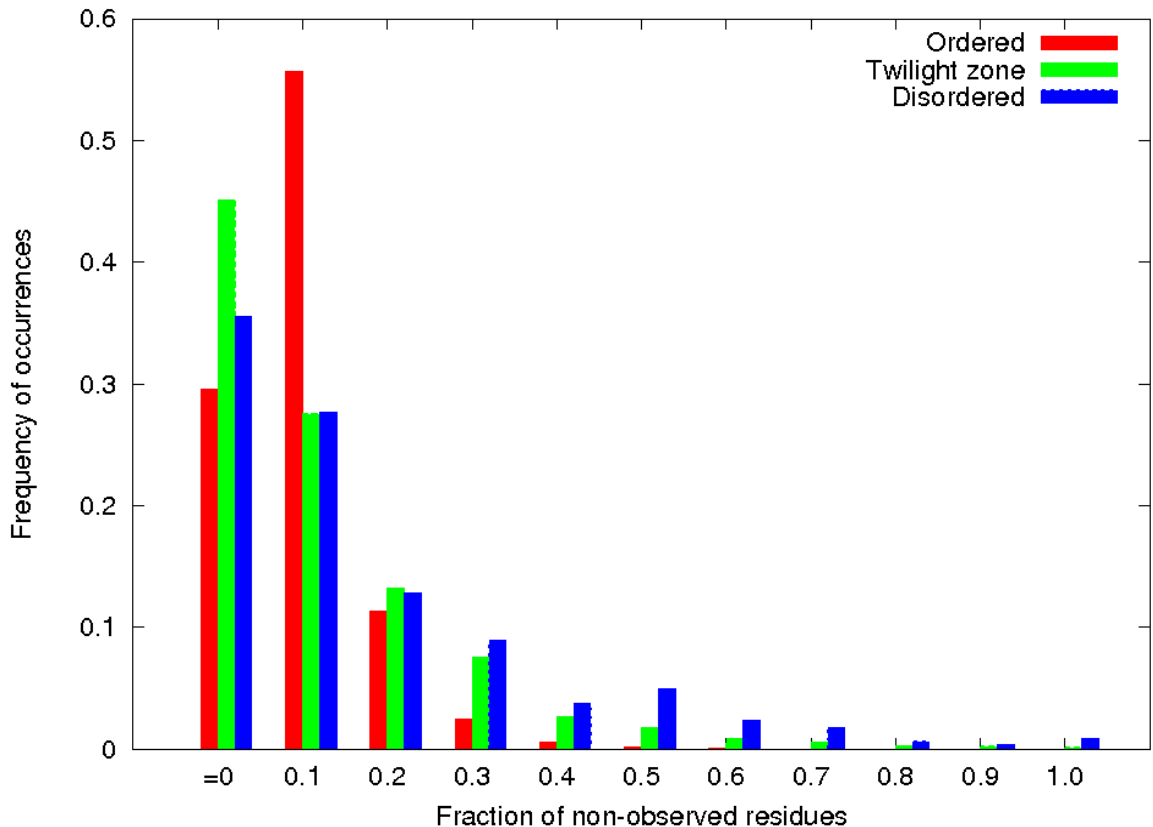

**Figure 5 – Distribution non-observed residues in proteins predicted as ordered, disordered and in the twilight zone**

Red bars refer to proteins with S > 3.24 (ordered), green bars refer to proteins belonging to the twilight zone, blue bars refer to proteins with S < -0.83 (disordered). The fraction of non-observed residues (horizontal axis) is the ratio between the number of non-observed residues and the number of the amino acids in a protein sequence. The fraction of non-observed residues has been divided in bins. The first bin refers to proteins without non-observed residues. The second bin refers to proteins with a fraction of non-observed residues between 0 and 10%, the third bin refers to proteins with a fraction of non-observed residues between 11% and 20% and so on. The frequency of occurrences (vertical axis) for each group of proteins is the number of proteins with a given fraction of non-observed residues divided by the number of proteins considered. A large fraction of proteins predicted as disordered and in the twilight zone do not have non-observed residues.

# Appendix

**Frequency of proteins predicted as natively unfolded by gVSL2 and mean pairwise energy $<E_C>$**

| ORGANISM | N. proteins | Predictions | | | |
|---|---|---|---|---|---|
| | | gVSL2 | | $<E_C>$ | |
| | | N. | % | N. | % |
| **ARCHAEA** | | | | | |
| *Aeropyrum pernix* | 1776 | 129 | 7.26 | 38 | 2.14 |
| *Archaeoglobus fulgidus* | 2411 | 134 | 5.56 | 36 | 1.49 |
| *Methanococcus jannaschii* | 1784 | 93 | 5.21 | 16 | 8.95 |
| *Pyrococcus abyssii* | 1787 | 76 | 4.25 | 18 | 1.01 |
| *Thermoplasma volcanium* | 1524 | 74 | 4.86 | 34 | 2.23 |
| | **9282** | **506** | **5.45** | **142** | **1.53** |
| **BACTERIA** | | | | | |
| *Agrobacterium tumefaciens C58* | 5358 | 454 | 8.47 | 452 | 8.44 |
| *Aquifex aeolicus* | 1557 | 93 | 5.97 | 18 | 11.56 |
| *Chlamydia pneumoniae* | 2756 | 336 | 12.19 | 169 | 6.13 |
| *Chlorobium tepidum TLS* | 2286 | 247 | 10.80 | 172 | 7.52 |
| *Escherichia coli K12* | 5336 | 392 | 7.35 | 343 | 6.43 |
| *Haemophilus influenzae Rd* | 31910 | 2357 | 7.39 | 1758 | 5.51 |
| *Mycobacterium tuberculosis H37Rv* | 60406 | 8917 | 14.76 | 9397 | 15.56 |
| *Neisseria meningitidis MC58* | 1901 | 126 | 6.63 | 132 | 6.94 |
| *Salmonella typhi* | 5355 | 386 | 7.21 | 355 | 6.63 |
| *Staphylococcus aureus COL* | 2680 | 265 | 9.89 | 205 | 7.65 |
| *Synechocystis species PCC 6803* | 3532 | 256 | 7.25 | 231 | 6.54 |
| *Thermotoga maritima* | 1891 | 125 | 6.61 | 35 | 1.85 |
| *Treponema pallidum* | 3495 | 317 | 9.07 | 198 | 5.67 |
| | **128463** | **14271** | **11.11** | **13465** | **10.48** |
| **EUKARYA** | | | | | |
| *Anopheles gambiae* | 15821 | 4430 | 28.00 | 3748 | 23.69 |
| *Arabidopsis thaliana* | 51094 | 15678 | 30.68 | 10788 | 21.11 |
| *Bos taurus* | 15658 | 4986 | 31.84 | 3774 | 24.10 |
| *Caenorhabditis elegans* | 23354 | 6717 | 28.76 | 5041 | 21.59 |
| *Drosophila melanogaster* | 34043 | 13117 | 38.53 | 10861 | 38.53 |
| *Gallus gallus* | 8392 | 2603 | 31.02 | 1988 | 23.69 |
| *Homo sapiens* | 95581 | 34749 | 36.36 | 28732 | 30.06 |
| *Mus musculus* | 65284 | 24519 | 37.56 | 18624 | 28.53 |
| *Oryza sativa* | 98035 | 43272 | 44.14 | 34890 | 35.59 |
| *Plasmodium falciparum* | 5356 | 1974 | 36.86 | 914 | 17.06 |
| *Saccharomyces cerevisiae* | 41117 | 11975 | 29.12 | 8612 | 20.95 |
| | **453735** | **164020** | **36.15** | **127972** | **28.20** |